
\tolerance=100000
\hyphenpenalty=1000
\raggedbottom
\font\twelvebf=cmbx12

\def\ainit{\hoffset=.0 truecm
            \voffset=1. truecm
            \hsize=17. truecm
            \vsize=23.5 truecm
            \baselineskip=11.pt
            \lineskip=0pt
            \lineskiplimit=0pt}
\def\pag{\pageno=2\footline={\hss\tenrm\folio\hss}}
\ainit
%
%

\def\mic{\,\mu{\rm m}}

\def\yr{\,{\rm yr}}

\def\Msol{\,{\rm m_\odot}}
\def\Lsol{\,{\rm L_\odot}}

\def\Msyr{\,{{\rm m_\odot}\,{\rm yr}^{-1}}}
%
%

\def\lsim{\,\lower2truept\hbox{${< \atop\hbox{\raise4truept\hbox{$\sim$}}}$}\,}
\def\gsim{\,\lower2truept\hbox{${> \atop\hbox{\raise4truept\hbox{$\sim$}}}$}\,}

%
%
\def\oneskip{{\vskip \baselineskip}}
\centerline{\null}
\nopagenumbers
\oneskip
\noindent
\centerline  {\twelvebf A model for the continuum energy distribution}
\centerline  {\twelvebf of the ultraluminous galaxy IRAS~F10214$+$4724}
\bigskip

\oneskip
\oneskip
\parindent=1truecm
\parskip 0pt

\centerline{P. Mazzei and G. De Zotti}

\oneskip
\noindent
{\it Osservatorio Astronomico, Vicolo dell'Osservatorio 5,
 I--35122 Padova, Italy}

\oneskip
\oneskip
\oneskip
\noindent
\settabs 2 \columns
\+ Address for correspondence: &Dr. Gianfranco De Zotti\cr
\+&Osservatorio Astronomico\cr
\+&Vicolo dell'Osservatorio, 5\cr
\+&I--35122 Padova (Italy)\cr
\+&e-mail: 39003::DEZOTTI\cr
\+&or DEZOTTI@ASTRPD.ASTRO.IT\cr

\oneskip
\noindent

\vfill\eject

\pag
\noindent
{\bf ABSTRACT}

\medskip
\noindent
If indeed early type galaxies used up most of their gas to form stars in a
time short compared to their collapse time and if a roughly constant fraction
of metals is locked up in dust grains, these galaxies may easily become
opaque to starlight and emit most of their luminosity in the far-IR. The
corresponding spectral energy distribution matches remarkably well the
observed continuum spectrum of the ultraluminous galaxy IRAS F$10214+4724$
from UV to sub-mm wavelengths, i.e. over almost four decades in frequency,
for a galactic age $\lsim 1\,$Gyr. The bolometric luminosity in this
model is $\simeq 2.7\times 10^{14}\Lsol$ ($H_0 =50\,\hbox{km}\,\hbox{s}^{-1}
\,\hbox{Mpc}^{-1}$, $\Omega =1$), i.e. somewhat lower than
implied by previous models. In the present framework, the bolometric
luminosity of the galaxy is expected to decrease by a factor $\gsim 30$
during the subsequent evolution.


\bigskip\noindent
{\bf 1\qquad INTRODUCTION}

\medskip\noindent
Optical searches for protogalaxies and, in general, for high redshifts galaxies
have been remarkably unsuccessful, in
spite of many years of efforts (Djorgovski, Thompson, \& Smith 1993;
Cowie, Songaila, \& Hu 1991; Colless et al. 1993), while a couple of
apparently genuine young galaxies have been discovered in
samples selected at longer wavelengths, i.e. in the radio (53W002,
at $z=2.390$; Windhorst et al. 1991) and in the far-IR (IRAS F$10214+4724$
at $z=2.287$; Rowan-Robinson et al. 1991). An additional actively star
forming galaxy at $z\simeq 2$ may have been detected as a damped Ly$\alpha$
absorber (Elston et al. 1991). In view of this fact and of the difficulties
facing the alternative merging scenario, at least as far as bright galaxies
are concerned (see Franceschini et al. 1993), it is worthwhile to investigate
the possible effect of dust on the optical visibility of early phases of
galaxy evolution (van den Berg 1990; Wang 1991; Kormendy \& Sanders 1992).

Mazzei et al. (1992, 1993) have worked out models for the
broad band photometric evolution of disk and spheroidal galaxies from UV to
millimetric wavelengths, taking into account in a self
consistent way the chemical evolution of the interstellar medium, the
evolution of the dust content, the corresponding extinction and
reradiation by dust, under the assumption that galaxies were closed
systems. Broadly speaking, disk galaxies correspond to the case
of a slow gas depletion, i.e. of a star formation rate (SFR) never much higher
than it is today, while spheroidal galaxies are thought to have used up
most of their gas to form stars in a time short compared to the collapse
time, so that their SFR should have been initially very high
(Sandage 1986). Thus, the spectral energy distribution of disk galaxies
should not have varied much during most of their lifetime; on the
contrary, spheroidal galaxies are expected to have been much brighter in their
early evolutionary phases. If the early metal enrichment was fast enough,
and the dust to gas ratio is roughly proportional to the metallicity,
the galaxy may easily become optically thick during these early gas
rich, very high luminosity phases.

Franceschini et al. (1993) have investigated
implications of these models for deep optical,
near-IR, far-IR and radio counts of galaxies and for the corresponding
redshift distributions, to conclude that models implying
an opaque early phase in the evolution of spheroidal galaxies can account
both for the lack of high redshift galaxies in deep optical surveys and
for the deep $60\mic$ IRAS counts and also for a good fraction of
sub-mJy radio sources detected in VLA surveys.

In this paper, we compare the continuum spectral energy distribution
predicted by such models with observational data on IRAS F$10214+4724$,
the putative protogalaxy for which the most extensive spectral coverage
is presently available (Rowan-Robinson et al. 1991; Soifer et al. 1991;
Clements et al. 1992; Downes et al. 1992; Lawrence et al. 1993).
In the general framework described above,
the observational evidence that this galaxy
is undergoing a giant starburst on the scale of the whole galaxy suggests
that it is a proto-elliptical (Rowan-Robinson et al. 1991; Elbaz et al. 1992).
In Section 2 we will then describe
the key features of our evolutionary models for spheroidal galaxies.
In Section 3 we will compare model predictions with observational data
and discuss some implications. In Section 4 we outline the main
conclusions. Throughout this paper we use $H_0 =50\,\hbox{km}\,\hbox{s}^{-1}
\,\hbox{Mpc}^{-1}$ and $\Omega =1$.

\bigskip\noindent
{\bf 2\qquad THE PHOTOMETRIC EVOLUTION MODEL}

\medskip\noindent
Our evolutionary synthesis model is described by Mazzei et al. (1993).
Here we recall only the basic ingredients.

The unreddened stellar energy distribution from UV to mid-IR (N band)
was obtained as follows.
The number of stars born at each galactic age and the gas metallicity were
computed by solving the standard equations governing chemical evolution.
Schmidt's (1959) parametrization
($\psi(t)=\psi_0 f_g(t)^n\, \Msol \yr^{-1}$ where $f_g$ is the gas fraction)
has been adopted for the star formation rate (SFR). As for the initial
mass function (IMF), we have considered both a Scalo's (1986) and a
Salpeter's (1955) form; in the latter case, different choices for the
lower mass limit have been explored. The isochrones by Bertelli et al. (1990),
extended by Mazzei (1988), have been exploited to get
the distribution of stars of each generation in the H-R diagram.

Dust extinction was estimated assuming a spherically symmetric King's (1966)
distribution of well mixed stars, gas and dust. The dust to gas ratio
was taken to be proportional to the gas metallicity $Z_g$.
The optical depth is thus proportional to $f_g(t)Z_g(t)$ and increases
rapidly with decreasing galactic age. In our baseline model we adopt
the coefficient of proportionality determined by Mazzei et al. (1993)
from their analysis of a sample of local early type galaxies.
We have also employed the slightly more elaborate model proposed by
Guiderdoni \&
Rocca-Volmerange (1987) whereby the dust to gas ratio is
proportional to $Z_g^s$ with $s=1.6$
for $\lambda > 2000\,\AA$ and $s=1.35$ at shorter wavelengths; the results
are basically similar.
The standard mean extinction curve for the interstellar
medium in our own Galaxy (Seaton 1979; Rieke \& Lebofsky 1985) was adopted.

The starlight absorbed by dust is reemitted in the far-IR. Its
spectral distribution
was modelled  following Xu \& De Zotti (1989),
i.e. taking into account the contributions of two components: warm dust,
located in regions of high radiation field intensity (i.e. in the
neighborhood of OB associations), and cold dust, heated by the
general interstellar radiation field. The warm to cold dust radiation
intensity was taken to be proportional to the star formation rate, as
described in Mazzei et al. (1993). Given the extreme star formation
activity of IRAS F$10214+4724$, the warm dust component is strongly
dominant.

Emission from circumstellar dust shells around OH/IR stars was also
taken into account. The total luminosity of such stars at each galactic
age is taken to be 5\% of the global luminosity of AGB stars with initial
masses in the appropriate range.

\bigskip\noindent
{\bf 3\qquad RESULTS AND DISCUSSION}

\medskip\noindent
If indeed a roughly constant fraction of metals present at any time
is locked into dust grains, the dust extinction in spheroidal galaxies
is bound to be large during the early evolutionary phases
when a substantial metal enriched interstellar medium was
present. If the gas metallicity increases fast enough, the galaxy becomes
opaque during the first 1 or 2 Gyrs of its life, so that most of the
luminosity comes out at far-IR wavelengths, as observed for
IRAS F$10214+4724$.

An example, corresponding to a power law index $n=0.5$ for the SFR and to a
Salpeter's IMF, is shown in Fig. 1. A very good fit of the data is obtained
for a galactic age $T \simeq 1\,$Gyr, implying a redshift of formation
$z_f \simeq 4$. The SFR at this age is $3\times 10^4\Msyr$ and the
average extinction is $A_B \simeq 4.6\,$mag.
This model has significantly less power in the IR than the best fit model by
Rowan-Robinson et al.
(1993; their model B) and, correspondingly, a significantly lower
bolometric luminosity ($L_{\hbox{bol}} \simeq 2.7\times 10^{14} \Lsol$).

The metallicity is predicted to be roughly solar, consistent with the
(still uncertain) observational indications (Downes et al. 1992; see,
however, Brown \& Vanden Bout 1992 for a significantly lower estimate).

Figure 1 illustrates the photometric evolution predicted by the model,
showing that the far-IR to optical luminosity ratio decreases very fast
with increasing galactic age, reaching the very low values typical of
present early type galaxies at an age of $\sim 15\,$Gyr.

It must be stressed that, in this general framework, the age of
the best fit model is not unambiguously determined. Quite generally,
however, we may conclude that it is unlikely to be substantially higher
than $T = 1\,$Gyr. This is because the luminosity of the UV branch decreases
as stellar populations get older making increasingly difficult to account for
the observed UV flux in the presence of the strong absorption required
to explain the far-IR .

On the other hand, it is possible to obtain
a fit comparable to that in Fig. 1 assuming a lower galactic age (see,
for example, Fig. 2). As shown by Fig. 1, the model gets bluer at
earlier times. Thus a simple recipe to obtain a fit with a lower $T$
is to assume a higher absorption during the earliest evolutionary phases;
as a result, the metallicity associated to the best fit model decreases,
if the other model parameters are kept unchanged.

Obviously, the metallicity at a given galactic
age is higher and higher as the IMF is more and more weighted
towards massive stars. The requirement that it does not
exceed the solar value then implies relatively low galactic ages.
For example, in the case of a Scalo's (1986) IMF, a solar metallicity
is reached at $T = 0.4\,$Gyr and in the case of a Salpeter's (1955) IMF
with a lower mass limit $m_l = 0.5\Msol$ at $T = 0.1\,$Gyr. The faster
increase of the metallicity naturally provides the stronger extinction
required to compensate for the more intense UV branch; only minor
adjustments of the optical depth are needed to get a good fit (see Fig. 2).

The total mass of the model galaxy also critically depends on the IMF. A
standard Salpeter's (1955) IMF with a lower mass limit $m_l = 0.01\Msol$
implies a total mass of $\simeq 5\times 10^{13}\Msol$. This estimate
however may need a substantial downward correction if indeed formation
of low mass stars is inhibited in starbursts. Increasing the lower
mass limit to $0.5\Msol$ would translate in a decrease of the estimated total
mass to $\simeq 7\times 10^{12}\Msol$.

It may be noted that a
galactic age  of $1\,$Gyr corresponds to the characteristic timescale
for gas consumption in elliptical galaxies; the subsequent evolution
is then be characterized by a rapid dimming (cf. Fig. 3)
amounting to a factor of more than 30 by the present time. Thus,
IRAS F10214$+4724$ could be the progenitor of a $z=0$ elliptical galaxy
with a bolometric luminosity of $\lsim 10^{13}\Lsol$.

\bigskip\noindent
{\bf 4\qquad CONCLUSIONS}

\medskip\noindent
Under the very simple assumption that the dust to gas ratio in galaxies
is always roughly proportional to the metallicity of the interstellar
medium, early type galaxies are expected to undergo an opaque phase
during their early evolution whenever the metal enrichment is sufficiently
rapid. We have shown that the ensuing spectral energy distribution
for a galactic age of $\lsim 1\,$Gyr matches
impressively well the observational data for the ultraluminous
IRAS F10214$+4724$
over almost four decades in wavelength, i.e. from UV to sub-mm, without
any need of invoking additional contributions from e.g. an active nucleus
and assuming the standard mean extinction curve for our own Galaxy.

While this galaxy is extraordinary as far as the luminosity is concerned,
it may well be that an optically thick early phase, corresponding to
the maximum bolometric luminosity, is a common property of early type
galaxies.
As discussed by Franceschini et al. (1993), a photometric
evolution of this kind
would simultaneously explain the lack of high redshift galaxies
in optically selected samples down to $B=25$ and the $60\mic$ counts of
galaxies in the deep IRAS survey.

\bigskip\noindent
{\bf ACKNOWLEDGEMENTS.}

\medskip\noindent
We are grateful to C. Xu who has worked out
the calculations of the diffuse dust emission spectrum. Work supported
in part by ASI and by the EEC programme ``Human Capital and Mobility''.

\vfill\eject

%
%
%
\def\aa #1 #2{{A\&A,}~{#1}, {#2}}
\def\aas #1 #2{{A\&AS,}~{#1}, {#2}}
\def\aar #1 #2{{A\&A Rev,}~{#1}, {#2}}
\def\advaa #1 #2{{Adv. Astron. Astrophys.,}~{#1}, {#2}}
\def\araa #1 #2{{ARA\&A,}~{#1}, {#2}}
\def\aj #1 #2{{AJ,}~{#1}, {#2}}
\def\alett #1 #2{{\it Astrophys. Lett.,}~{#1}, {#2} }
\def\apj #1 #2{{ApJ,}~{#1}, {#2}}
\def\apjs #1 #2{{ApJS,}~{#1}, {#2}}
\def\aplc #1 #2{{Ap. Lett. Comm.,}~{#1}, {#2}}
\def\apss #1 #2{{Ap\&SS,}~{#1}, {#2}}
\def\astrnach #1 #2{{Astron. Nach.,}~{#1}, {#2}}
\def\azh #1 #2{{AZh,}~{#1}, {#2}}

\def\baas #1 #2{{\it Bull. Am. astr. Soc.,}~{#1}, {#2} }

\def\ca #1 #2{{Comm. Astrophys.,}~{#1}, {#2}}
\def\cpc #1 #2{{Comput. Phys. Comm.,}~{#1}, {#2}}
\def\fcp #1 #2{{\it Fundam. Cosmic Phys.,}~{#1}, {#2} }
\def\jmp #1 #2{{J. Math. Phys.,}~{#1}, {#2}}
\def\jqsrt #1 #2{{J. Quant. Spectrosc. Rad. Transf.,}~{#1}, {#2}}
\def\memsait #1 #2{{\it Memorie Soc. astr. ital.,}~{#1}, {#2} }
\def\mnras #1 #2{{MNRAS,}~{#1}, {#2}}
\def\qjras #1 #2{{\it Q. Jl R. astr. Soc.,}~{#1}, {#2} }
\def\nat #1 #2{{Nature,}~{#1}, {#2}}
\def\pasj #1 #2{{PASJ,}~{#1}, {#2}}
\def\pasp #1 #2{{\it PASP,}~{#1}, {#2}}

\def\physl #1 #2{{\it Phys.Lett.,}~{#1}, #2}
\def\physrep #1 #2{{Phys. Rep.,}~{#1}, #2}
\def\physscri #1 #2{{Phys. Scripta,}~{#1}, #2}

\def\physreva #1 #2{{Phys. Rev. A,}~{#1}, {#2}}
\def\physrevb #1 #2{{Phys. Rev. B,}~{#1}, {#2}}
\def\physrevd #1 #2{{Phys. Rev. D,}~{#1}, {#2}}
\def\physrevl #1 #2{{Phys. Rev. Lett.,}~{#1}, {#2}}
\def\pl #1 #2{{\it Phys. Lett.,}~{#1}, {#2} }

\def\prsl #1 #2{{\it Proc. R. Soc. London Ser. A,}~{#1}, {#2} }

\def\ptp #1 #2{{Prog. Theor. Phys.,}~{#1}, {#2}}
\def\ptps #1 #2{{Prog. Theor. Phys. Suppl.,}~{#1}, {#2}}
\def\rmp #1 #2{{Rev. Mod. Phys.,}~{#1}, {#2}}
\def\rpp #1 #2{{Rep. Progr. Phys.,}~{#1}, {#2}}
\def\sci #1 #2 {{Science,}~{#1}, {#2}}
\def\sovastr #1 #2{{Sov. Astr.,}~{#1}, {#2}}
\def\sovastrl #1 #2{{Sov.Astr. Lett.,}~{#1}, #2}

\def\ssr #1 #2{{\it Space Sci. Rev.,}~{#1}, {#2} }
\def\va #1 #2{{\it Vistas in Astronomy,}~{#1}, {#2} }

\def\book #1 {{\it ``{#1}'',\ }}


\def\ref{\noindent\hangindent=20pt\hangafter=1}

\oneskip
\centerline {\bf References}
\oneskip
\parindent=0pt
\parskip=0pt

\ref
 Bertelli G., Betto R., Bressan A., Chiosi C., Nasi E., Vallenari A.,
1990, \aas 85 845

\ref
Brown R.L., Vanden Bout P.A., 1992, \apj 397 L11

\ref
Clements D.L., Rowan-Robinson M., Lawrence A., Broadhurst T., McMahon R.,
1992, \mnras 256 35P

\ref
Colless M., Ellis R., Taylor K., Hook R., 1993, \mnras 216 19

\ref
Cowie L., Songaila A., Hu E.M., 1991, \nat 354 460

\ref
Djorgovski S., Thompson D., Smith J.D., 1993, in B. Rocca-Volmerange,
M. Dennefeld, B. Guiderdoni, J. Tran Thanh Van, eds., proc. 8th IAP
workshop ``First Light in the Universe'', Ed. Fronti\`eres, Gif sur Yvette

\ref
Downes D., Radford S.J.E., Greve A., Thum, C., 1992, \apj 398 L25

\ref
Elbaz D., Arnaud M., Cass\'e M., Mirabel I.F., Prantzos N., Vangioni-Flam
E., 1992, \aa  265 L29

\ref
Elston R., Bechtold J., Lowenthal J., Rieke M., 1991, \apj 373 L39

\ref
Franceschini A., Mazzei P., De Zotti G., Danese L., 1993,  ApJS subm.

\ref
Guiderdoni B., Rocca-Volmerange B., 1987, \aa 186 1

\ref
King I., 1966, \aj 71 64

\ref
Kormendy J., Sanders D.B., 1992, \apj 390 L53

\ref
Lawrence A. et al., 1993, \mnras 260 28

\ref
Mazzei P., 1988, Ph. D. thesis, Intern. School for Advanced Studies, Trieste,
Italy.

\ref
Mazzei P., Xu C., De Zotti G., 1992, \aa  256 45

\ref
Mazzei P., De Zotti G., Xu C., 1993,  ApJ  subm.

\ref
Rieke G.H., Lebofsky M.J., 1985, \apj 288 618

\ref
Rowan-Robinson M. et al. 1991, \nat 351 719

\ref
Rowan-Robinson M. et al. 1993, \mnras 261 513

\ref
Salpeter E.E., 1955, \apj  121 161

\ref
Sandage A., 1986, \aa 161 89

\ref
Scalo, J.M., 1986, \fcp 11 1

\ref
Schmidt M., 1959, \apj 129 243

\ref
Seaton M.J., 1979, \mnras 187 73P

\ref
Soifer B.T., et al. 1991, \apj 381 L55

\ref
van den Berg S., 1990, \pasp 102 503

\ref
Wang B., 1991, \apj 374 465

\ref
Windhorst R.A., Burstein D., Mathis D.F., Neuschaefer L.W., Bertola
F., Buson L.M., Koo D.C., Matthews K., Berthal P.D., Chambers K.C.,
1991, \apj 380 362

\ref
Xu C., De Zotti G., 1989, \aa 225 12

\vfill\eject

\bigskip
\rm{FIGURE CAPTIONS}
\medskip
{\bf {Fig. 1.}} Spectral energy distribution for several galactic ages
for a Salpeter's IMF with a lower mass limit $m_l=0.01\,\Msol$,
a star formation rate proportional to
$f_g^{0.5}$ (see Sect. 2) and a gas to dust
ratio proportional to metallicity ($s=1$).
Data are from Rowan-Robinson et al. (1993) and Downes et al. (1992).
The best fit corresponds to a galactic age, $T$, equal to $1\,$Gyr.

\medskip
{\bf {Fig. 2.}}
Spectral energy distribution
at galactic ages of 0.1--0.4$\,$Gyr for the same model as in Fig. 1,
except for a larger lower mass limit ($m_l=0.5\,\Msol$); also, the coefficient
of proportionality for the optical depth (see Sect. 2)
has been increased by a factor 1.5.

\medskip
{\bf {Fig. 3.}} Bolometric luminosity of our baseline model as a function of
galactic age, assuming $\Omega=1\,$ (see text).

\bye